\newcommand{\CLASSINPUTtoptextmargin}{19mm}%
\newcommand{\CLASSINPUTbottomtextmargin}{43mm}%
\newcommand{\CLASSINPUTinnersidemargin}{12.9mm}%
\newcommand{\CLASSINPUToutersidemargin}{12.9mm}%
\documentclass[conference,10pt,a4paper]{IEEEtran}

\makeatletter

\def\@maketitle{\newpage
\bgroup\par\addvspace{0.5\baselineskip}\centering%
\ifCLASSOPTIONtechnote
   {\bfseries\large\@IEEEcompsoconly{\sffamily}\@title\par}\vskip 1.3em{\lineskip .5em\@IEEEcompsoconly{\sffamily}\@author
   \@IEEEspecialpapernotice\par{\@IEEEcompsoconly{\vskip 1.5em\relax
   \@IEEEtitleabstractindextextbox{\@IEEEtitleabstractindextext}\par
   \hfill\@IEEEcompsocdiamondline\hfill\hbox{}\par}}}\relax
\else
   \vskip0.2em{\EuMWtitlesize\ifCLASSOPTIONtransmag\bfseries\LARGE\fi\@IEEEcompsoconly{\sffamily}\@IEEEcompsocconfonly{\normalfont\normalsize\vskip 2\@IEEEnormalsizeunitybaselineskip
   \bfseries\Large}\@title\par}\vskip1.0em\par
   \ifCLASSOPTIONconference%
      {\@IEEEspecialpapernotice\mbox{}\vskip\@IEEEauthorblockconfadjspace%
       \mbox{}\hfill\begin{@IEEEauthorhalign}\@author\end{@IEEEauthorhalign}\hfill\mbox{}\par}\relax
   \else
      \ifCLASSOPTIONpeerreviewca
         {\@IEEEcompsoconly{\sffamily}\@IEEEspecialpapernotice\mbox{}\vskip\@IEEEauthorblockconfadjspace%
          \mbox{}\hfill\begin{@IEEEauthorhalign}\@author\end{@IEEEauthorhalign}\hfill\mbox{}\par
          {\@IEEEcompsoconly{\vskip 1.5em\relax
           \@IEEEtitleabstractindextextbox{\@IEEEtitleabstractindextext}\par\hfill
           \@IEEEcompsocdiamondline\hfill\hbox{}\par}}}\relax
      \else
         \ifCLASSOPTIONtransmag
           {\@IEEEspecialpapernotice\mbox{}\vskip\@IEEEauthorblockconfadjspace%
            \mbox{}\hfill\begin{@IEEEauthorhalign}\@author\end{@IEEEauthorhalign}\hfill\mbox{}\par
           {\vspace{0.5\baselineskip}\relax\@IEEEtitleabstractindextextbox{\@IEEEtitleabstractindextext}\vspace{-1\baselineskip}\par}}\relax
         \else
           {\lineskip.5em\@IEEEcompsoconly{\sffamily}\sublargesize\@author\@IEEEspecialpapernotice\par
           {\@IEEEcompsoconly{\vskip 1.5em\relax
            \@IEEEtitleabstractindextextbox{\@IEEEtitleabstractindextext}\par\hfill
            \@IEEEcompsocdiamondline\hfill\hbox{}\par}}}\relax
         \fi
      \fi
   \fi
\fi\par\addvspace{0.0\baselineskip}\egroup}

\def\EuMWtitlesize{\@setfontsize{\EuMWtitlesize}{24}{24pt}}
\def\EuMWauthorsize{\@setfontsize{\EuMWauthorsize}{11}{11pt}}
\def\EuMWaffilsize{\@setfontsize{\EuMWaffilsize}{10}{10pt}}
\def\EuMWcaptionsize{\@setfontsize{\EuMWcaptionsize}{9}{10pt}}
\def\EuMWbibsize{\@setfontsize{\EuMWbibsize}{8}{10pt}}

\def\@IEEEauthorblockNstyle{\EuMWauthorsize\@IEEEcompsocnotconfonly{\sffamily}\@IEEEcompsocconfonly{\large}}
\def\@IEEEauthorblockAstyle{\EuMWaffilsize\@IEEEcompsocnotconfonly{\sffamily}\@IEEEcompsocconfonly{\itshape}\@IEEEcompsocconfonly{\large}}
\def\@IEEEauthordefaulttextstyle{\EuMWauthorsize\@IEEEcompsocnotconfonly{\sffamily}\sublargesize}

\def\thebibliography#1{\section*{\refname}%
    \addcontentsline{toc}{section}{\refname}%
    \EuMWbibsize\@IEEEcompsocconfonly{\small}\vskip 0.3\baselineskip plus 0.1\baselineskip minus 0.1\baselineskip
    \list{\@biblabel{\@arabic\c@enumiv}}%
    {\settowidth\labelwidth{\@biblabel{#1}}%
    \leftmargin\labelwidth
    \advance\leftmargin\labelsep\relax
    \itemsep \IEEEbibitemsep\relax
    \usecounter{enumiv}%
    \let\p@enumiv\@empty
    \renewcommand\theenumiv{\@arabic\c@enumiv}}%
    \let\@IEEElatexbibitem\bibitem%
    \def\bibitem{\@IEEEbibitemprefix\@IEEElatexbibitem}%
\def\newblock{\hskip .11em plus .33em minus .07em}%
\ifCLASSOPTIONtechnote\sloppy\clubpenalty4000\widowpenalty4000\interlinepenalty100%
\else\sloppy\clubpenalty4000\widowpenalty4000\interlinepenalty500\fi%
    \sfcode`\.=1000\relax}

\long\def\@makecaption#1#2{%
\ifx\@captype\@IEEEtablestring%
\par\@IEEEtabletopskipstrut
\else
\@IEEEfigurecaptionsepspace
\fi
\setbox\@tempboxa\hbox{\normalfont\footnotesize {#1.}\nobreakspace\nobreakspace #2}%
\ifdim \wd\@tempboxa >\hsize%
\setbox\@tempboxa\hbox{\normalfont\footnotesize {#1.}\nobreakspace\nobreakspace}%
\parbox[t]{\hsize}{\normalfont\footnotesize\noindent\unhbox\@tempboxa#2}%
\else
\ifCLASSOPTIONconference \hbox to\hsize{\normalfont\footnotesize\hfil\box\@tempboxa\hfil}%
\else \hbox to\hsize{\normalfont\footnotesize\box\@tempboxa\hfil}%
\fi\fi
\ifx\@captype\@IEEEtablestring%
\@IEEEtablecaptionsepspace
\else
\fi}

\newlength\tablecaptiontotableskip
\newlength\figuretocaptionskip
\def\@IEEEfigurecaptionsepspace{\vskip\figuretocaptionskip\relax}%
\def\@IEEEtablecaptionsepspace{\vskip\tablecaptiontotableskip\relax}%

\def\abstract{\normalfont%
\@IEEEabskeysecsize\bfseries\textit{\abstractname}\,\bfseries\textit{---}\,%
\@IEEEgobbleleadPARNLSP}%

\def\IEEEkeywords{\normalfont%
\@IEEEabskeysecsize\bfseries\textit{\IEEEkeywordsname}\,\bfseries\textit{---}\,%
\@IEEEgobbleleadPARNLSP}%
\def\endIEEEkeywords{\relax\vspace{0.67ex}%
\par\if@twocolumn\else\endquotation\fi%
\normalsize\normalfont}%

\DeclareRobustCommand*{\EuMWauthorrefmark}[1]{\raisebox{0pt}[0pt][0pt]{\textsuperscript{\footnotesize{#1}}}}%
\def\@IEEEauthorblockNtopspace{0ex}
\def\@IEEEauthorblockAtopspace{1mm}
\def\tablename{Table}
\def\thetable{\arabic{table}}
\def\IEEEkeywordsname{Keywords}
\def\subsubsection{\@startsection{subsubsection}{3}{\z@}{1.5ex plus 1.5ex minus 0.5ex}%
{0.7ex plus .5ex minus 0ex}{\normalfont\normalsize\itshape}}%
\newlength{\CPheadmatchindent}%
\def\@seccntformat#1{\hbox to\CPheadmatchindent{\csname the#1dis\endcsname}\hskip 0.1em \relax}
\IEEEilabelindentA \parindent
\IEEEilabelindent \IEEEilabelindentA
\IEEEelabelindent \parindent
\IEEEdlabelindent \parindent
\IEEElabelindent \parindent
\makeatother

\usepackage{paralist}
\usepackage[utf8]{inputenc}
\usepackage[backend=biber,style=ieee,maxnames=3,minnames=2]{biblatex}
\usepackage{amsmath,amssymb,amsfonts}
\usepackage{textcomp}
\usepackage{mathtools}
\usepackage{adjustbox}
\usepackage{balance}
\usepackage{bm}
\usepackage{xcolor}
\def\BibTeX{{\rm B\kern-.05em{\sc i\kern-.025em b}\kern-.08em
    T\kern-.1667em\lower.7ex\hbox{E}\kern-.125emX}}

\usepackage[hidelinks]{hyperref}
\usepackage{cleveref}
\usepackage{tabularx}
\usepackage{float}

\usepackage{todonotes}
\usepackage{numprint}     
\usepackage{makecell}   
\usepackage{xcolor,colortbl}
\usepackage{enumitem}            
\usepackage{subcaption}
\usepackage{wrapfig}
\usepackage{booktabs}
\usepackage{graphics}
\usepackage{graphicx}
\usepackage{makecell}

\usepackage{amsthm}
\usepackage{amsmath, amssymb}
\usepackage{placeins}

\usepackage{tikz}

\usepackage[acronym]{glossaries}

\newcommand{\egc}{e.\,g., }

\crefname{figure}{Fig.}{Fig.}

\begin{document}

\raggedbottom

\title{6G Integrated Sensing and Communication: From Vision to Realization}

\author{
    \IEEEauthorblockN{
        Thorsten Wild\EuMWauthorrefmark{\#},
        Artjom Grudnitsky\EuMWauthorrefmark{\#},
        Silvio Mandelli\EuMWauthorrefmark{\#},
        Marcus Henninger\EuMWauthorrefmark{\#},
        Junqing Guan\EuMWauthorrefmark{\#},
        Frank Schaich\EuMWauthorrefmark{\#}
        }

	\IEEEauthorblockA{
	\EuMWauthorrefmark{\#}Nokia Bell Labs Stuttgart, 70469 Stuttgart \\
	E-mail: \{firstname.lastname\}@nokia-bell-labs.com \\
 \it{Submitted to the convened focus session ``Joint Communication and Radar Sensing - a step towards 6G"}
 }
 }

\maketitle

\newacronym{agv}{AGV}{automated guided vehicle}
\newacronym{bf}{BF}{beamforming}
\newacronym{bs}{BS}{base station}
\newacronym{dft}{DFT}{Discrete Fourier Transform}
\newacronym{dl}{DL}{downlink}
\newacronym{e2e}{E2E}{end-to-end}
\newacronym{fmcw}{FMCW}{frequency modulated continuous wave}
\newacronym{isac}{ISAC}{Integrated Sensing and Communication}
\newacronym{kf}{KF}{Kalman Filter}
\newacronym{ofdm}{OFDM}{orthogonal frequency-division multiplexing}
\newacronym{papr}{PAPR}{peak-to-average power ratio}
\newacronym{poc}{PoC}{proof-of-concept}
\newacronym{qos}{QoS}{quality of service}
\newacronym{scfde}{SC-FDE}{single carrier frequency domain equalization}
\newacronym{uav}{UAV}{unmanned aerial vehicle}
\newacronym{ue}{UE}{user equipment}
\newacronym{ul}{UL}{uplink}
\newacronym{semf}{SeMF}{sensing management function}
\newacronym{snr}{SNR}{signal-to-noise ratio}
\newacronym{tdd}{TDD}{time division duplex}
\newacronym{tx}{TX}{transmit}

\begin{abstract}
Integrated sensing and communications (ISAC) will be deployed into cellular communication systems possibly already with 5G-A and surely in 6G.
This paper discusses ISAC use cases, key technology building blocks for system design with solutions and open research questions. Furthermore, we introduce our \gls{poc} based on commercially available 5G communications hardware at mm-Wave frequencies, with sensing-specific algorithmic extensions. This new ISAC \gls{poc} can perform jointly high data-rate communications and OFDM radar sensing in the same frequency band. Initial pedestrian detection results are shown, indicating the practicability of ISAC in future cellular networks. The results also indicate our achievable sensing range and provide hints to the achievable range estimation accuracy, based on the stability of the \gls{poc} system communications hardware.
\end{abstract}

%

\section{Introduction}\label{sec:intro}
6G cellular communication is predicted to be commercially ready in 2030. A key element of its vision is forming around the connection of different worlds \cite{viswanathan2020}. This is a representation of the physical/biological world in software,  thus digital twinning. Digital twins can play an important role especially in the industrial sector to enhance productivity, safety and security. In each wireless generation, larger bandwidths have been used, denser networks, more processing capabilities and a larger number of antennas. Under these trends with ubiquitously deployed cellular communication systems in place, the integration of RF sensing is becoming more and more attractive to extend the network service capabilities and to enable the digital twins with the necessary context information from the environment. The 6G network will become the sensor \cite{de2021convergent, wei2022toward, wild2021joint}, building on cellular system strengths such as interference management, wide area coverage and tightly controlled and scheduled operation. 5G systems already have active localization capabilities \cite{henninger2022probabilistic}, where positioning reference signals are transmitted from \glspl{ue} or \glspl{bs}, also denoted as gNBs. Furthermore, for 5G systems, 3GPP standardization has identified possible use cases for \gls{isac} \cite{3gpp}, which would allow to detect also objects which are not connected to the network. There are large public funded projects, e.g. \cite{hexa_x}, which address the collaborative work towards \gls{isac} technologies and system design for the coming years.

In this paper we discuss 6G \gls{isac} starting from spectrum and key use cases. Technology building blocks, available solutions and research challenges are addressed and finally we describe a novel real-time \gls{isac} \gls{poc}, which builds upon commercially available 5G communications hardware in combination with dedicated sensing processing elements. We make use of the communications split processing architecture interfaces for sensing purposes and \gls{ofdm} radar processing algorithms \cite{Braun2014_OFDM_Radar}, enhanced by clutter removal and tracking algorithms. In our \gls{poc}, sensing can be carried out with down to 0\% sensing overhead on top of running communication services. We show example results for human detection hinting to achievable range and range estimation stability and accuracy.
\section{Basic Overview overview on 6G ISAC}\label{sec:overview}
\begin{figure*}[!t]
    \begin{center}
        \includegraphics[width=0.95\textwidth]{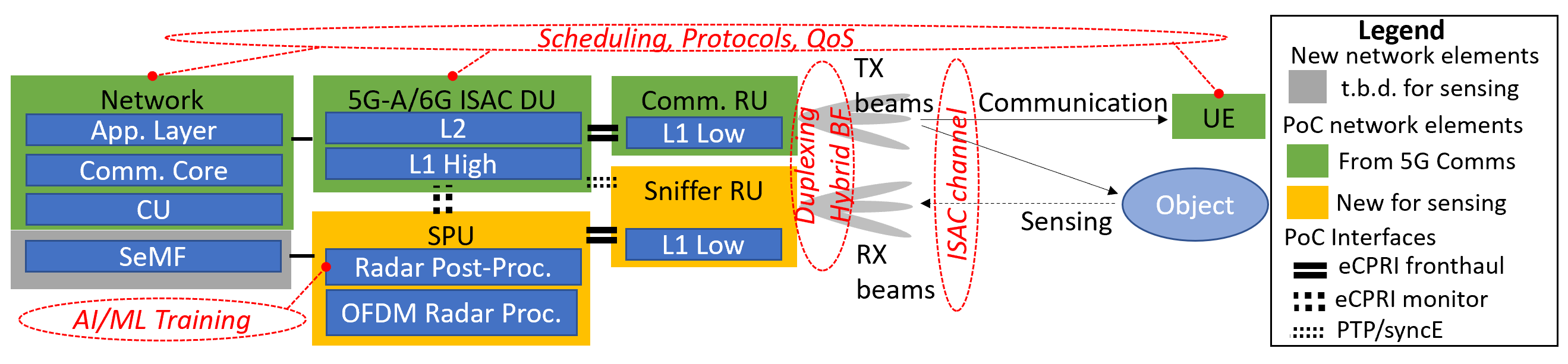}
    \end{center}
\caption{System overview incl. our \gls{poc} processing blocks and associated research challenges (red) – mono-static example.}
  \label{fig:sys_overview}
\vspace{-5mm}
\end{figure*}
\gls{isac} makes very efficient use of spectrum as, due to its integrated nature, the same bands with the same carriers can be used for both communication and sensing purposes, and the multiplexing capabilities of cellular systems (\egc due to \gls{ofdm} waveforms) can efficiently trade-off service qualities both for communications as well as the new to-be-introduced sensing service qualities. The spectrum choice dominates the system capabilities. 5G capacity layers are in the so-called FR1 (frequency range 1) bands below 5 GHz and use bandwidths up to 100 MHz. FR2 in the millimeter frequency ranges starting at around 24 GHz offer bandwidths up to 800 MHz and are more attractive for sensing due to the higher range resolution. For 6G, the main capacity layer, the ``Golden Band", is expected to be in 6-14 GHz and may offer up to 400 MHz of bandwidth, making it attractive for wide-area sensing. In order to re-use 5G site-grids, we can expect massive MIMO \glspl{bs} with around 1024 antenna elements (offsetting coverage drawbacks relative to FR1), thus we can also leverage large antenna apertures for angular resolution. Refarming and spectrum sharing of lower FR1 bands may also happen but will come with limited available bandwidths. This means that we can see the Golden Band and FR2 millimeter wave spectrum as the strongest sensing introduction candidates in 6G. Sub-THz technologies with carrier frequencies \textgreater100 GHz are also highly discussed for 6G and would offer huge bandwidths. Looking at the commercial introduction of 5G, the FR1 band was dominating over the FR2 band due to better coverage. If 6G deployments will be driven along the commercial needs of communication systems first, we can assume that also sub-THz sensing may initially not play the strongest role in early 6G.

We can expect that the key \gls{isac} use cases will be driven and stimulated by the frequency-dependent deployment scenarios. Micro-cellular deployments mainly rely on FR2 (and alternatively/later sub-THz) and will be used indoor and in hot-spot areas. In factory floor environments, pedestrians, passive factory objects or other moving objects (such as robots, \glspl{agv} and fork lifts) can be detected . This could enhance safety in industrial environments. Macro-cellular deployments will dominate outdoors, where mainly the Golden Bands are of key interest for sensing. Traffic monitoring, highway intrusion detection, roadside safety, drone/\gls{uav} intrusion detection or tracking are all examples of possible use cases. Cellular networks can provide the infrastructure-based sensing where local sensors would be limited by not being able to ``look around the corner" from the device point of view.

Regarding waveforms for \gls{isac}, the aforementioned \gls{ofdm} is fully established in 4G and 5G communication systems and is also gaining more momentum in the radar world. Therefore, \gls{ofdm} is an attractive option to multiplex communication and sensing services in the same carrier band. Its only drawback is the \gls{papr}, so in case of \egc higher carrier frequencies, communication systems may use \gls{scfde} as an alternative, which could leverage the same frequency domain sensing processing as \gls{ofdm}. Other waveform alternatives (such as time-multiplexing of communication waveforms with \gls{fmcw}) are of course not ruled out and can also be studied.
\section{Technology Building Blocks}
\label{sec:building_blocks}


Sensing in cellular systems can be achieved by different architectures. 
If the sensing processing resides in the \gls{bs}, the system has lighter signaling overhead and less demands concerning \gls{ue} complexity and battery consumption. Sensing transmitters located at the \gls{bs} offer a larger range  (due to higher transmit powers and advantageous placement). 
%
Mono-static sensing requires an almost full duplex capability, where the self-interference can be reduced by a slightly separated receive array in combination with regular communications \gls{tdd} operation, as we will discuss in~\ref{sec:Poc_Realization}. 
Moreover, mono-static solutions offer easier use of data-aided sensing, which has the opportunity to avoid sensing overhead in case an active communication transmission is ongoing in the sensing direction of interest.
Bi-static sensing can avoid this issue, but requires receiving sensing entities to mute transmissions, which reduces the temporal availability of the \gls{isac} system, and thus increases the additional overhead for sensing on top of the communication system. 
Using both \gls{ue}s and \gls{bs} is a further bi-static sensing alternative which requires the availability of a \gls{ue} at the location of interest.

In our \gls{poc} we focus on the network-based mono-static \gls{isac}. 
Fig.~\ref{fig:sys_overview} displays a simplified \gls{e2e} system overview (and further more our used \gls{poc} interfaces and building blocks as described in ~\ref{sec:Poc_Realization} are marked) with the \gls{bs} split architecture with radio unit (RU), distributed unit (DU) and central unit (CU). It remains a research and standardization challenge to come up with accepted \gls{isac} channel models which serve well for simulating both communication and sensing performance offering some level of statistical abstraction. 
Agreed communication channel models for performance benchmarks are statistically based; they so far fail to support the required sensing features of full geometrical consistency in combination with a radar cross section parameter. A statistical model for \gls{isac} fixing those issues with the option of ray-tracing-assisted cluster placement could be an attractive way forward for 3GPP and thus needs to be studied in Rel. 19, where Ch.~8 in~\cite{3gppchan} could serve as a starting point.

To implement \gls{isac}, radio front-end design must take into account new challenges and review solutions for ones already well investigated.
Full duplex requires to deal with self-interference. We circumvent this issue by using a separated \emph{Sniffer} RU for reception of the reflections. With appropriate array design and self-interference compensation mechanisms the communication array and the Sniffer may be integrated as subpanels into a single RU (or alternatively, based on the split processing interfaces described in \ref{sec:Poc_Realization} be placed further apart for bi-/multi-static operation). Note that during \gls{ul}, both arrays could be used to improve diversity and \gls{snr} for the communication mode. Another issue arises from the presence of analog and hybrid \gls{bf} in FR2 and even Golden Band frequencies. To get a radar image of the environment with pre-defined grid of beams codebooks implemented in wireless products, the angular domain must be properly sampled and interpolated. SARA \cite{sara2022} addresses this and defines how lossless reconstruction of the array's angular response can be achieved by uniform sampling in the \gls{dft} beamspace of the system's sum co-array and its corresponding interpolation.


Scheduling of both communication as well as sensing signals in the same carrier poses new design questions and opportunities. The proportional fair metric in communication schedulers \cite{pf2014} should be expanded towards integrating (i)~legacy communications and (ii)~to-be-defined sensing \gls{qos}. The \gls{e2e} network architecture of 5G systems needs to be expanded by new network functionalities dedicated to sensing, which we could denote \gls{semf}. This central function will be responsible for aggregating sensing information from \glspl{bs} and \gls{ue}s, and providing high-level resource allocation management to cells involved in sensing operations based on sensing \gls{qos}.

With 6G as AI-native air interface, AI/ML for sensing can leverage available AI-hardware acceleration and is attractive to be used for tasks such as object detection and classification, both locally in the \gls{bs} or centrally in the \gls{semf}.
\section{Proof of Concept Architecture}\label{sec:Poc_Realization}
A major design goal of the \gls{isac} architecture is to re-use as much of the communication hardware as possible for sensing purposes.
The \gls{ofdm} radar approach~\cite{Braun2014_OFDM_Radar} uses the frequency domain \gls{tx} reference signal and frequency domain reflected signal as inputs into the radar processing chain.
Looking at possible functional splits for 5G~\cite{Larsen_2019_Functional_Splits}, using Option 7-2 (FFT/iFFT done in RU, frequency-domain IQ transported on fronthaul) allows use of the same type of RU for both communication and sensing, and using the fronthaul between DU and RU as the input to \gls{ofdm} radar processing.

\Cref{fig:sys_overview} shows also the architecture of our \gls{poc}.
The building blocks in green are based on a commercial 5G FR2 communication system we use as a foundation.
To enable \gls{isac} we extend this system with (i) a dedicated RU called \emph{Sniffer} of the same type as the communication RU and (ii) a server called Sensing Processing Unit (\emph{SPU}) running our real-time \gls{ofdm} radar processing software, highlighted in orange.
While the gNB RU is operating in TDD mode (4:1 DL/UL ratio), the Sniffer is always in \gls{ul}-mode, thus it receives both reflected signals and uplink signals from \glspl{ue}.

The communication setup uses eCPRI~\cite{eCPRI_2_0_Spec} over optical fiber (25 GE) to carry Sync-, Management-, Control- and User-planes.
A fiber-optical splitter on the DU$\rightarrow$RU fiber allows us access to the \gls{tx} waveform and \gls{bf} information used by the communication setup.
The Sniffer is directly controlled by the SPU, which switches beams and requests the radio resources to be received over Control-plane, and receives IQ-data of the reflected signal via User-plane. 
Both RUs are connected via fiber to a common sync source (5G system module), synchronizing both RUs via PTP~\cite{PTPv2_spec} and SyncE~\cite{SyncE_spec}, resulting in a synchronization jitter of $\approx$40ps between gNB RU and Sniffer.
The SPU server hardware uses 25 GE network cards to allow fronthaul access of gNB RU and Sniffer and synchronization of the SPU to the same PTP source (5G system module) as the RUs.
A CUDA capable GPU is used for \gls{ofdm} radar processing and visualization.
\begin{figure*}
  \begin{subfigure}[t]{0.22\linewidth}
    \begin{center}
      \includegraphics[width=1.0\linewidth,clip,trim={0mm 20mm 30mm 20mm}]{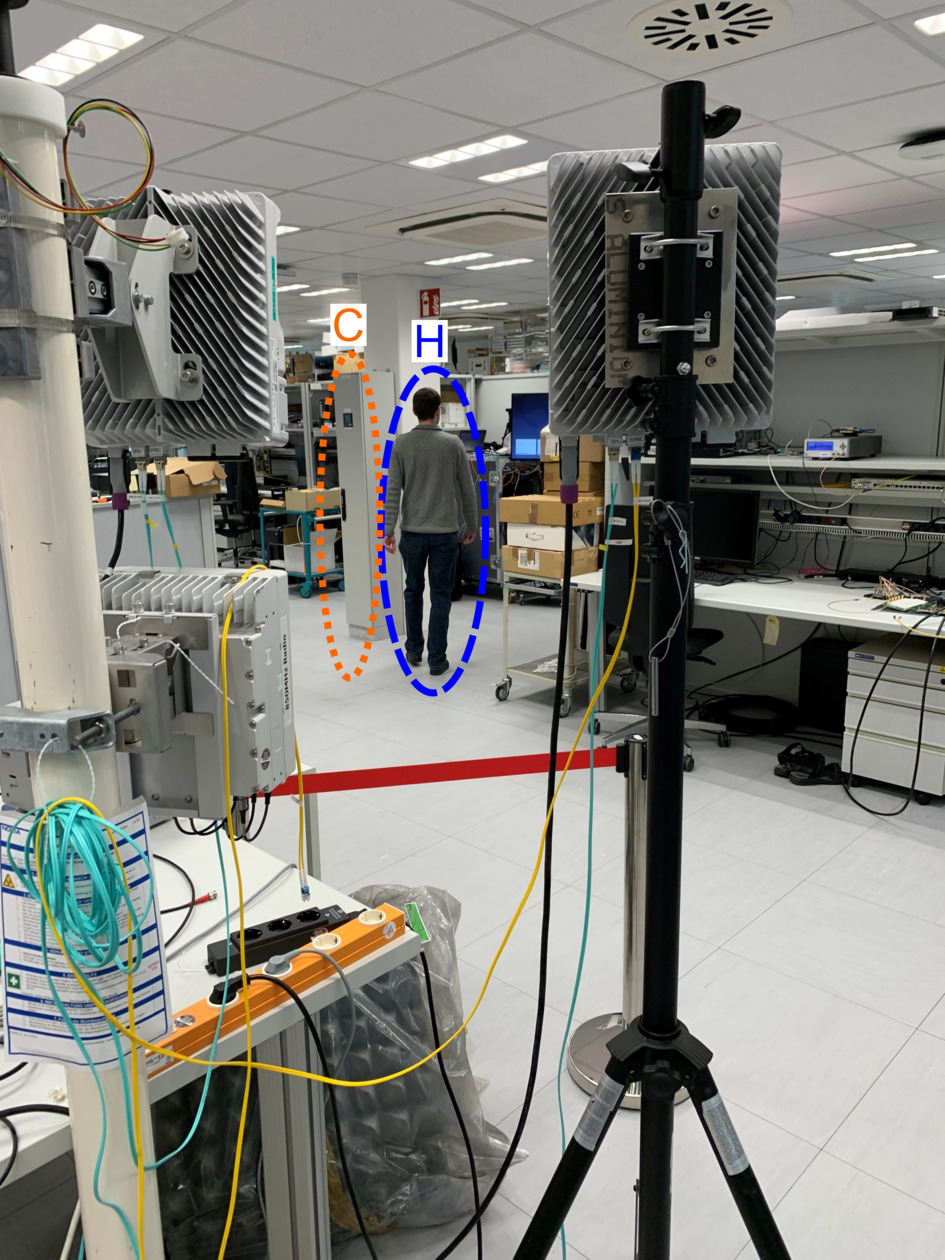}
    \end{center}
    \caption{Photo from the RUs (gNB RU top left, Sniffer top right) in the direction of the beam.}
    \label{fig:res_lab_walking_photo}
  \end{subfigure}\enspace%
  \begin{subfigure}[t]{0.35\linewidth}
    \begin{center}
      \includegraphics[width=1.0\linewidth]{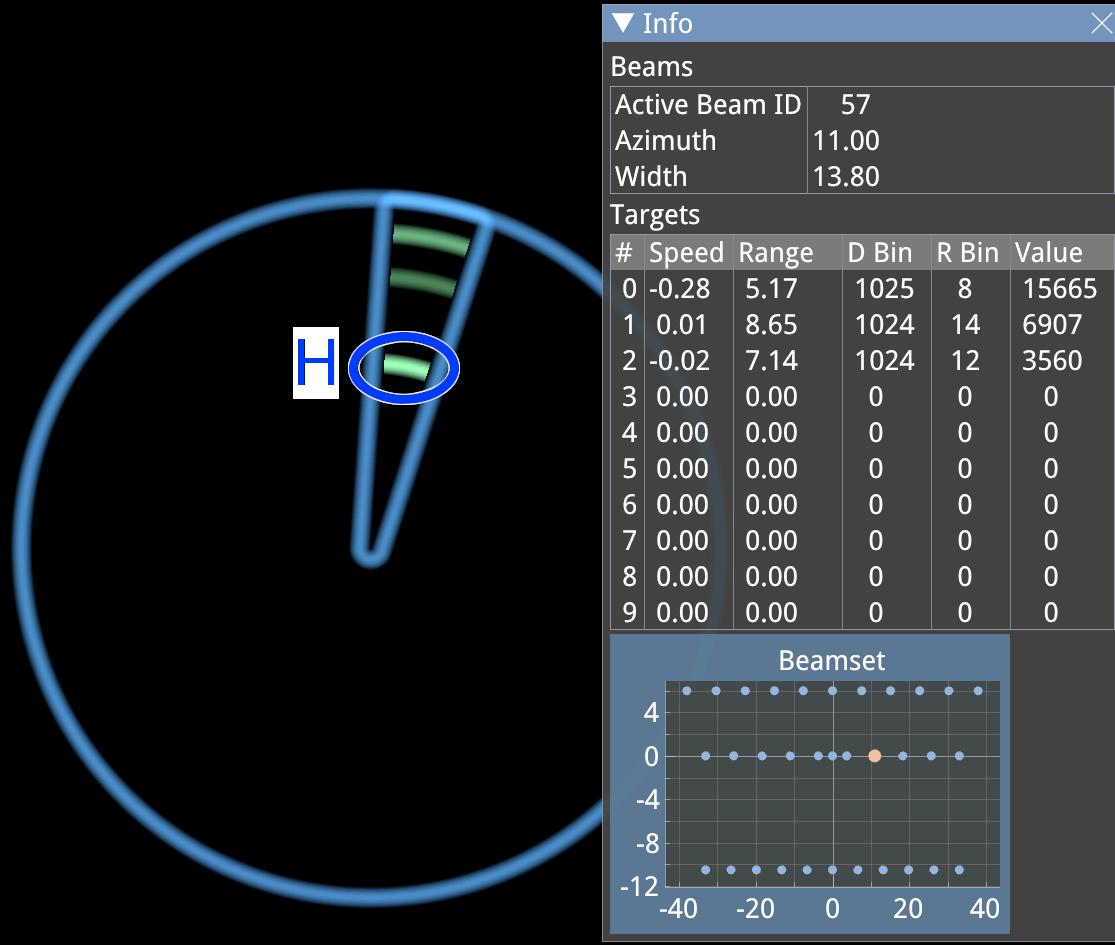}
    \end{center}
    \caption{Realtime visualization by our SPU software after \emph{Peak Extraction and Interpolation}. The target corresponding to the human \textbf{H} is highlighted.}
    \label{fig:res_lab_walking_spu_core}
  \end{subfigure}\enspace%
  \begin{subfigure}[t]{0.43\linewidth}
    \begin{center}
      \includegraphics[width=1.0\linewidth,clip,trim={0mm 10mm 0mm 0mm}]{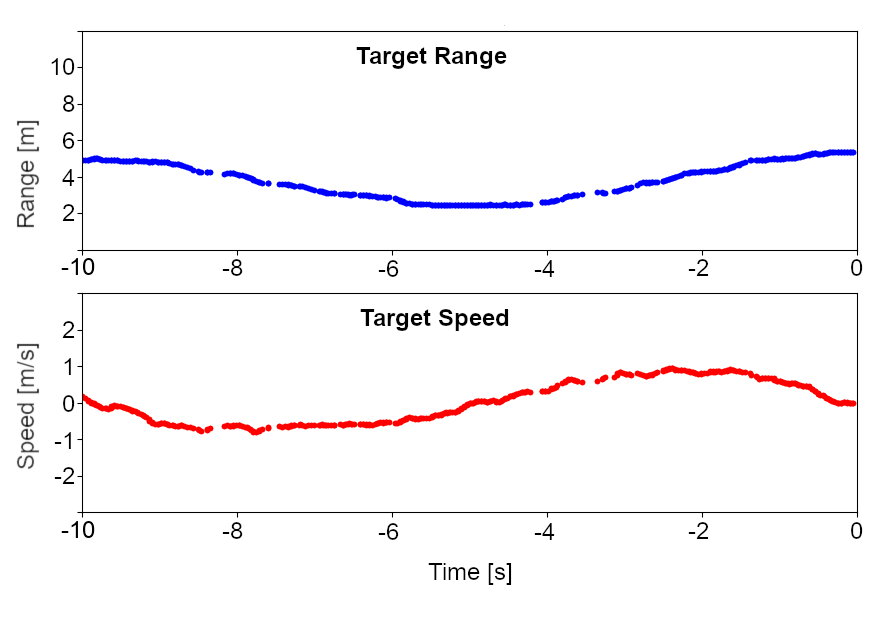}
    \end{center}
    \caption{Range and speed of the human over the past 10 seconds after \emph{Target Tracking}.}
    \label{fig:res_lab_walking_hos}
  \end{subfigure}
  \caption{Testing our ICAS PoC using a ``Walking Human'' scenario, with person \textbf{H} walking towards and away from the RUs. Clutter removal (main clutter \textbf{C} highlighted) was active.}
  \label{fig:res_lab_walking_scenario}
  \vspace{-5mm}
\end{figure*}

The SPU processing chain comprises the following steps:
\begin{compactitem}
\item \emph{Reference and Reflected Signal Reception}. Receive and preprocess eCPRI packets from gNB RU and Sniffer.
Each radio frame is an $N\times M$ matrix, where $N$ is the number of subcarriers (SC) in each symbol, and $M$ is the number of symbols in the radio frame.
\item \emph{Channel Computation}. Perform resource-element-wise division of the reflected signal by the reference signal.
\item \emph{Clutter Removal}. To isolate targets, it is crucial to remove signal contributions resulting from reflections from objects not of interest for the sensing task.
We record the channel of the reference scenario during system calibration and subtract it from the reflected signal at runtime.
\item \emph{Range/Doppler Periodogram Computation}. Perform $N$ FFTs on the channel, followed by $M$ iFFTs on the result.
The squared magnitude of each element of the result of the FFTs and iFFTs operations are computed, resulting in a 2D-periodogram, containing Doppler and range information for the processed radio frame.
\item \emph{Peak Extraction and Interpolation}. Elements in the periodogram that exceed a configurable threshold are potential targets.
The remaining elements usually form clusters in the range and Doppler domain, caused by the same object.
We extract the local maxima of these clusters and use them to construct the target list (sorted by element value, which corresponds to received power).
\item \emph{Target Tracking}. Detected targets are optionally tracked by means of a simple \gls{kf}.
\end{compactitem}

\section{First Results}
\label{sec:Results}

Using the system described in the previous section in a lab, we track a walking human.
The system operates at a center frequency of 27.6 GHz with 200 MHz bandwidth and a $\mu=3$ numerology (120 kHz SC spacing, 1120 symbols per frame).
In this experiment we use a test signal sent by the gNB RU (instead of a UE connection), which uses all SCs within the DL part of the radio frame modulated with 16-QAM.
The beam ($14^\circ$ of half-power horizontal beam width) is kept static during the test.
Both RUs are mounted at a height of ca. 1.9m, with a distance of $\approx$40cm between gNB RU and Sniffer.

\Cref{fig:res_lab_walking_scenario} shows (a) a photo of the scenario, (b) the SPU real-time visualization ``radar view'', with the person corresponding to the green arc within the beam cone, and (c) the distance from the RU and speed over the past 10 seconds of the strongest reflector (the person in this experiment), filtered by the  \gls{kf}.

\begin{figure}
    \begin{center}
      \includegraphics[width=1.0\linewidth,clip,trim={3mm 3mm 5mm 4mm}]{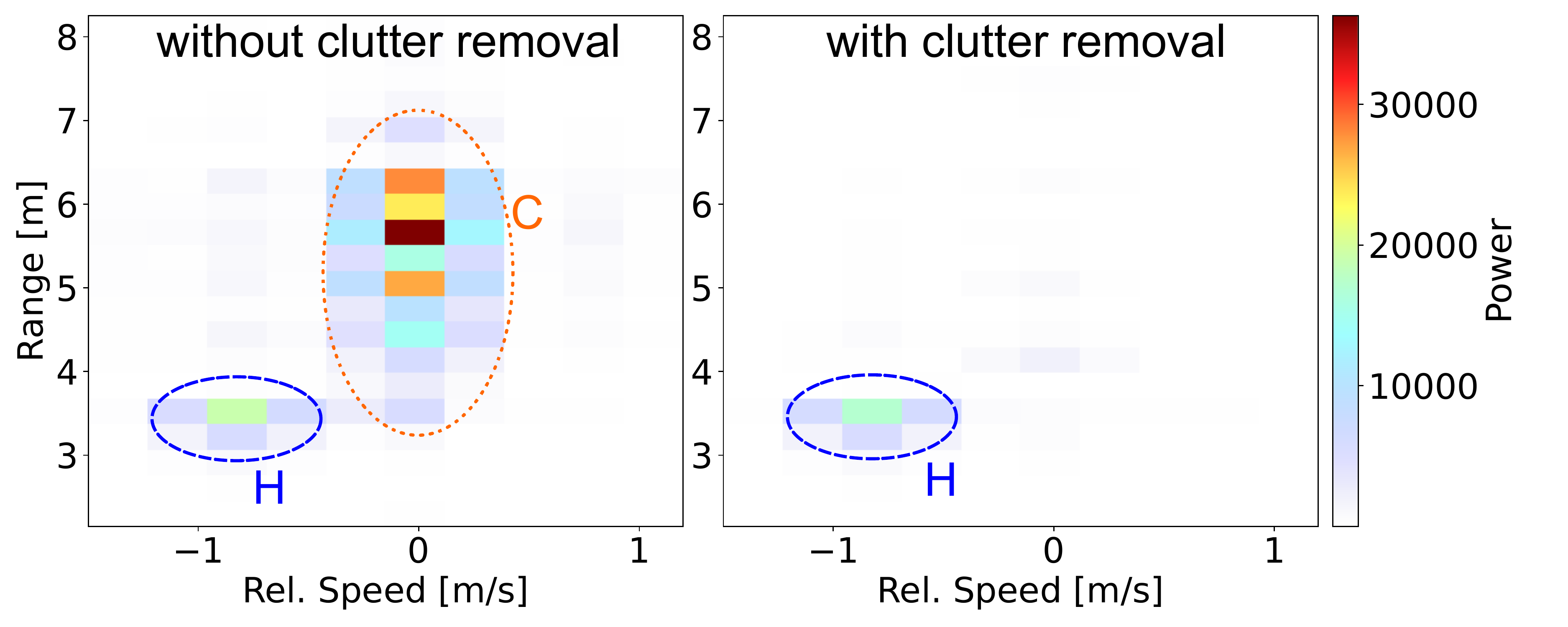}
    \end{center}
    \caption{Periodograms of same radio frame with and without clutter removal. Clutter \textbf{C} produces much stronger reflection than human \textbf{H}.}
    \label{fig:res_pgram_clutter}
  \vspace{-5mm}
\end{figure}

\Cref{fig:res_pgram_clutter} shows the effect of clutter removal -- indoor environments often contain objects (such as a metal cabinet and support pillar \textbf{C}) that cause stronger reflections than a human.
With clutter removal, the effects of these reflectors are removed, resulting in a periodogram with only the target \textbf{H} clearly visible.
This makes algorithms which use the results of OFDM radar as inputs (\egc Target Tracking) more robust.

In our first setup evaluations, we experienced a range standard deviation of $\sigma_{H} = 5$ mm for a human target at $r_H^0 = 3$ m range, while a strong clutter component at ca. 5.5~m range yields $\sigma_{C} = 1$ mm, before any tracking algorithms are considered. Given the estimated \gls{snr} $\gamma_H(r_H^0) = 10^{6.885}$ at $r_H^0$ range for humans, and assuming free space path loss exponent $\eta = 2$, a rough assumption of the \gls{snr} experienced at a generic range $r$ for human sensing with our setup can be written as
\begin{equation}
\gamma_H(r) = \gamma_H(r_H^0) \cdot (r_H^0/r)^{2 \eta} \;.
\end{equation}
Assuming a minimum \gls{snr} (after combining gains) $\gamma_0 = 10^{1.7}$ in order to obtain a reasonable spectral estimation performance~\cite{MUSIC_collapse}, one can determine the achievable range by our setup on human sensing as
\begin{equation}
r_H^* = \sqrt[2\eta]{\frac{\gamma_H(r_H^0)}{\gamma_0}} \cdot r_H^0 \approx 60\; m \; .
\end{equation}

\section{Conclusion}\label{sec:Conclusion}

The background for system design for \gls{isac} in cellular 5G-A and 6G systems has been discussed. 
Technical building blocks with solution options and research challenges have been addressed. 
Then, a novel \gls{poc} implementation with results has been presented, which shows that existing millimeter-wave 5G hardware can provide a good basis for adding sensing capabilities on top and that pedestrians can be detected well and become separated from clutter with appropriate algorithms.

\section*{Acknowledgments}
This work was developed within the KOMSENS-6G project, partly funded by the German Ministry of Education and Research under grant 16KISK112K.


\FloatBarrier
\setlength{\biblabelsep}{8pt}
\printbibliography

\end{document}